\documentclass[a4paper, fleqn, usenatbib]{mnras}

\usepackage{graphicx}

\usepackage{amsmath}
\usepackage{amssymb}
\usepackage{color}

\title[A wind-driving disc in HL Tau]{
  A wind-driving disc model for the mm-wavelength polarization structure of
  HL Tau}

\author[T. Matsakos, P. Tzeferacos and A. K\"onigl]{
  Titos Matsakos,\thanks{E-mail: titos.matsakos@uchicago.edu}
  Petros Tzeferacos
  and Arieh K\"onigl \\
  Department of Astronomy \& Astrophysics, The University of Chicago, 5640
  S.~Ellis Ave, Chicago, IL 60637, USA
}

\date{Accepted XXX. Received YYY; in original form ZZZ}
\pubyear{2016}

\begin{document}
\label{firstpage}
\pagerange{\pageref{firstpage}--\pageref{lastpage}}
\maketitle

\begin{abstract}
The recent advent of spatially resolved mm- and cm-wavelength polarimetry in
protostellar accretion discs could help clarify the role of magnetic fields in
the angular momentum transport in these systems.
The best case to date is that of HL~Tau, where the inability to produce a good
fit to the 1.25-mm data with a combination of vertical and azimuthal magnetic
field components was interpreted as implying that centrifugally driven winds
(CDWs) are probably not a significant transport mechanism on the $\sim 10^2\,$au
scale probed by the observations.
Using synthetic polarization maps of heuristic single-field-component discs and
of a post-processed simulation of a wind-driving disc, we demonstrate that a
much better fit to the data can be obtained if the radial field component, a
hallmark of the CDW mechanism, dominates in the polarized emission region.
A similar inference was previously made in modelling the far-infrared
polarization map of the pc-scale dust ring in the Galactic centre.
To reconcile this interpretation with theoretical models of protostellar discs,
which indicate that the wind is launched from a comparatively high elevation
above the mid-plane, we propose that most of the polarized emission originates
-- with a high ($\ga 10$\%) intrinsic degree of polarization -- in small
($\la 0.1\,$mm) grains that remain suspended above the mid-plane, and that the
bulk of the mm-wavelength emission is produced -- with low intrinsic
polarization -- by larger grains that have settled to the mid-plane.
\end{abstract}


\section{Introduction}
\label{sec:intro}

Magnetic fields are thought to be crucial to the transport of angular momentum
in accretion discs, but the precise nature of their effect is still an open
question.
The two leading scenarios that have been debated in the literature both require
an ordered, large-scale field (characterized by a thermal-to-magnetic pressure
ratio $\beta$) to be present.
In one picture, the transport takes place in the plane of the disc either
through MHD turbulence that is induced by the magnetorotational instability
\citep[MRI; e.g.][]{Balbus11} or, if the field is strong enough, through
large-scale correlations in a laminar flow \citep[e.g.][]{Simon+13}; in this
case $\beta_0\gg 1$ (where the subscript 0 denotes the mid-plane).
In the alternative scenario, angular momentum is transported vertically through
the disc surfaces in the form of a centrifugally driven wind \citep[CDW; e.g.][]
{KoniglSalmeron11}; it is efficient only if $\beta_0$ is not $\gg 1$.
It was, however, realized early on \citep{Li96, Wardle97} that, in the case of
weakly ionized discs where the field--matter coupling near the mid-plane is
weak, a CDW can dominate the angular momentum transport even if $\beta_0 \gg 1$.
In such discs, the wind is launched from a comparatively high elevation (a few
thermal-pressure scale heights above the mid-plane), where ionizing radiation
from an external source can penetrate the disc and ensure good coupling.
It was subsequently also realized \citep{Salmeron+07} that both mechanisms can
in principle operate at the same radial location in the disc -- with radial
transport dominating closer to the mid-plane and vertical transport becoming
important at a height where the density (and, correspondingly, $\beta$) is
sufficiently low for the magnetic field to suppress the MRI and drive a wind.
Recent numerical simulations of $\beta_0\gg 1$ protoplanetary discs have
indicated that high-elevation CDWs arise naturally in such systems -- even if
the mid-plane coupling remains adequate -- and that they can potentially play a
role in the angular momentum transport all along the disc, with radial transport
contributing to varying degrees at lower disc elevations \citep[e.g.][]
{BaiStone13, Bai13, Simon+13, Lesur+14, Bai15, Gressel+15, Simon+15}.

The continuum radiation of molecular discs is dominated by the thermal emission
of dust.
This emission will be polarized if non-spherical dust grains become aligned with
respect to the local magnetic field.
Various mechanisms can potentially effect such an alignment, with radiative
torque \citep[e.g.][]{DraineWeingartner96} being a favoured possibility for
protoplanetary discs.
If the polarized emission can be spatially resolved, it may be used to probe the
magnetic field structure in such discs \citep[e.g.][]{ChoLazarian07} and
potentially shed light on the angular momentum transport process.

Molecular accretion discs are found not just in protostars but also -- on much
larger scales -- in galactic nuclei, both active \citep[e.g.][]{Tristram+14} and
quiescent.
In fact, the first polarimetric observations of such a disc were carried out at
far-infrared ($\sim$$100\,\micron$) wavelengths for the pc-scale dust ring in
the Galactic centre \citep{Hildebrand+90, Hildebrand+93}.
The data for this disc, which is observed at a fairly large inclination angle
($i \approx 70^\circ$), could be interpreted in terms of radial ($r$) and
azimuthal ($\phi$) magnetic field components that are of comparable magnitude
but opposite signs, and a relatively weak vertical ($z$) field component,
consistent with the expectations for a weakly ionized disc in which a CDW
dominates the angular momentum transport \citep{WardleKonigl90}.

It has now also become possible to resolve the magnetic field structure of
protoplanetary discs through interferometry at sub-mm to cm wavelengths.
Data are already published for the young (Class-I/II) source HL~Tau
\citep[$1.25\,$mm;][]{Stephens+14} and for three Class-0 protostars:
IRAS~16293-2242B \citep[$878\,\micron$;][]{Rao+14}, L1527 \citep[$1.3\,$mm;][]
{Segura-Cox+15} and IRAS~4A \citep[$8.1\,$mm and $1.03\,$cm;][]{Cox+15}.
In contrast to HL~Tau, the even younger Class-0 sources are deeply embedded
within an extended envelope that can contribute significantly to the observed
emission \cite[e.g.][]{Looney+00}.
Furthermore, the gravitational collapse of such envelopes may create magnetized
pseudo-disc structures that are not genuine (i.e. rotationally supported) discs
\citep[e.g.][]{HennebelleCiardi09}.
In the unambiguous case of the HL~Tau disc ($i\approx47^\circ$),
\citet{Stephens+14} attempted to fit the data with a combination of azimuthal
and vertical field components.
They found that, while a pure-$B_\phi$ field did not fit the data well, the
addition of a $B_z$ component only made the fit worse, which led them to the
conclusion that a CDW is probably not the dominant angular momentum transport
mechanism on the $\sim$80-au scale probed by their observations.
In this connection it is worth noting that the protoplanetary disc simulations
discussed in \citet{Simon+13, Simon+15} and \citet{Bai15} indicate that, at
large ($\ga 30\;$au) radii, the region near the mid-plane could be either
turbulent or laminar (depending on the magnetic field strength and the
ionization structure) but that in either case the local field would be dominated
by the azimuthal component (essentially because of the strong shear of the
Keplerian velocity field; cf. \citealt{EardleyLightman75}).
Another noteworthy fact is that in HL~Tau there is evidence not just for a
central high-velocity jet \citep[e.g.][]{Mundt+90, Movsessian+12} but also for
an extended, lower-velocity disc wind \citep[e.g.][]{Takami+07,
LumbrerasZapata14, Klaassen+16}, which is consistent with general expectations
for actively accreting protostars \citep[e.g.][]{Ferreira+06}.

In this paper we demonstrate that a considerably better fit to the HL~Tau
polarization data can be obtained if one also includes the radial field
component -- which was neglected in the \citet{Stephens+14} analysis -- in the
modelling.
Indeed, we find that the best fit corresponds to
$|B_r| \ga |B_{\phi}| \gg |B_z|$, similar to the result previously obtained for
the Galactic-centre disc.
(Note, however, that the apparent polarization structures of these two sources
are not the same, which can be understood from the difference in their values of
the viewing angle $i$.)
Thus, in contradistinction to the inference of \citet{Stephens+14}, we find that
the polarization data not only do not negate the presence of a CDW in this disc
but that they actually support it.
The prominence of the radial field component in wind-driving discs is consistent
with the requirement that the ratio $B_r/B_z$ at the disc's surface exceed a
certain minimum value ($=1/\sqrt{3}$ for a cold, Keplerian disc;
\citealt{BlandfordPayne82}) in order for a CDW to be launched.

We complement the modelling approach of \citet{WardleKonigl90}, who employed a
combination of semi-analytic and analytic methods, by using a numerical
simulation of a disc with a generic magnetic diffusivity to derive the expected
polarization properties of a wind-driving disc.
These results are presented in Section~\ref{sec:model} and confronted with the
HL~Tau data in Section~\ref{sec:results}; in the latter Section we also show
synthetic maps for single-component field geometries (pure $B_r$, $B_\phi$ and
$B_z$) to help guide the analysis.
In Section~\ref{sec:discussion} we discuss our findings in relation to models of
the dust distribution in protoplanetary discs and to an alternative
interpretation of the data in terms of dust scattering \citep{Kataoka+15,
Kataoka+16, Yang+16a, Yang+16b}; we also comment on the applicability of this
model to the observed Class-0 sources. We summarize in
Section~\ref{sec:conclusion}.

\section{Numerical model and synthetic polarization maps}
  \label{sec:model}

Our disc model represents the final configuration of a numerical simulation that
evolves simultaneously the disc and its outflow, following the approach of
\cite{Tzeferacos+09, Tzeferacos+13}.
We set up a protoplanetary accretion disc that is threaded by a large-scale
poloidal magnetic field and is characterized by an isotropic, alpha-type
resistivity and an alpha-type viscosity. We integrate the axisymmetric MHD
equations (evolving all three vector components in the $r$--$z$ plane) using the
PLUTO code\footnote{
Freely available at: \texttt{http://plutocode.ph.unito.it/}}
\citep{Mignone+07, Mignone+12}.
This model assumes good field--matter coupling throughout the disc,
corresponding to the Elsasser number $\Lambda$ \citep[e.g.][]{KoniglSalmeron11}
being $\ga 1$  at all heights, as well as $\beta_0 \ga 1$.
However, in real protostellar discs, the mid-plane Elsasser number ($\Lambda_0$)
might drop below~1 and $\beta_0$ could be $\gg 1$ on the spatial scales of
interest \citep[e.g.][]{Simon+13, Simon+15, Bai15}.
Furthermore, the magnetic diffusivity on these scales would be dominated by
ambipolar diffusion (with a possible contribution from the Hall term; e.g.,
\citealt{WardleKonigl93, Bai15}) and would thus be anisotropic.
Nevertheless, as we further elaborate in Section~\ref{sec:discussion}, our model
should capture the key properties of a wind-driving disc that are germane to its
observed polarization structure.
 
For reference, we adopt the following set of parameters: sound-to-Keplerian
speed ratio $e = 0.1$, magnetization $\mu = \beta_0^{-1} = 0.4$, magnetic field
inclination parameter $m = 0.3$, magnetic diffusivity parameter
$\alpha_m = 0.01$ (with the parameter $\chi_m$ being 1, corresponding to full
isotropy), and magnetic Prandtl number (the ratio of viscosity to magnetic
diffusivity) $\mathcal{P}_m = 1$ (see \citealt{Tzeferacos+13} for further
details).
For these parameter choices, $\Lambda_0 = (2\mu)^{1/2}/\alpha_m \approx 89$.
The final outcome of the simulation consists of a self-consistent steady-state
disc--wind system, which we reconstruct in 3D for post-processing.

\begin{figure}
  \includegraphics[width=\columnwidth]{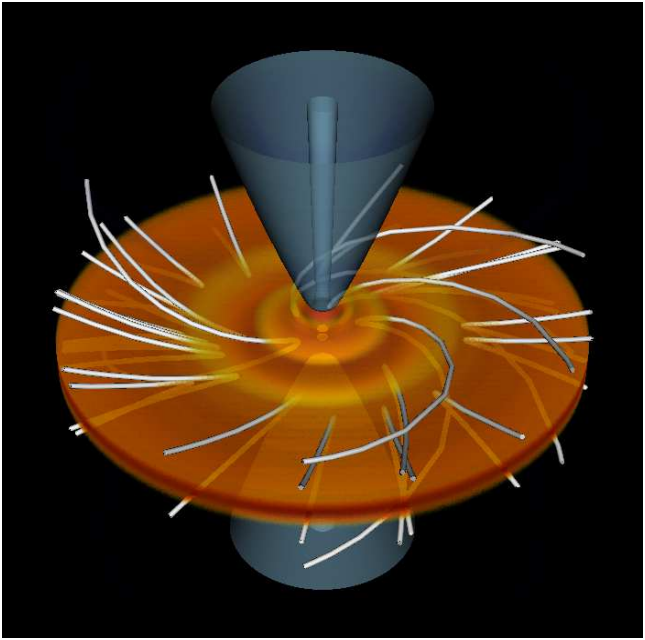}
  \caption{
    Composite 3D rendering of the disc--wind structure, viewed at an angle of
    $45^\circ$.
    The orange shading depicts the density distribution in the disc, and the
    cyan iso-surfaces highlight the central (highest velocity) portion of the
    wind.
    (The wind is driven from all disc radii, but its outer portion is not
    shown.)
    Also plotted (in white) is a random sampling of magnetic field lines that
    thread the disc.
    \label{fig:3Ddisk}}
\end{figure}
\begin{figure*}
  \includegraphics[width=\textwidth]{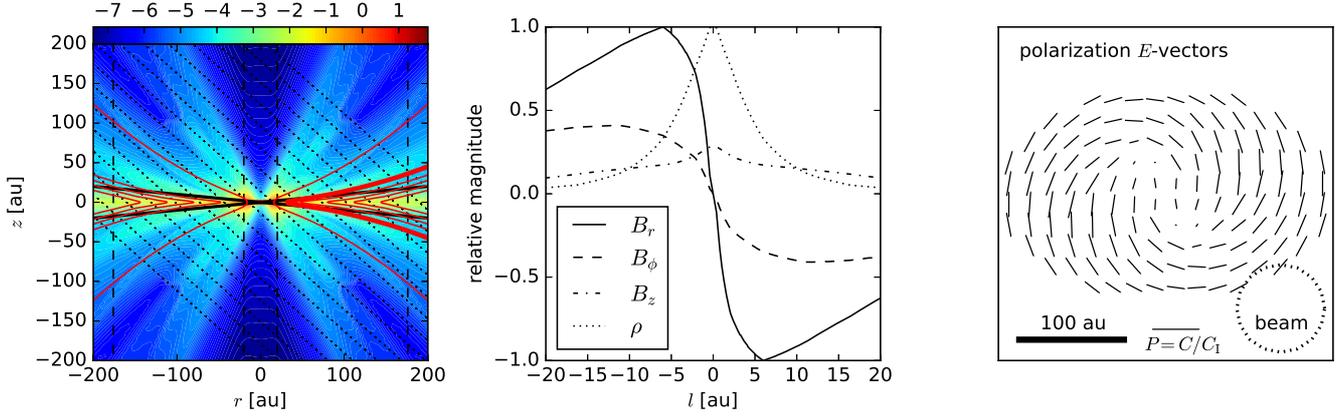}
  \caption{
    Structure and polarization properties of the simulated disc--wind model.
    Left panel: vertical slice showing the logarithmic density distribution
    (colour contours in code units) and the poloidal magnetic field lines (in
    red).
    The solid black lines approximate the nominal surfaces of the disc (the
    locations where the radial velocity component changes sign), the vertical
    dashed lines delimit the region that is included in the analysis, and the
    slanted dotted lines represent the directions along which the Stokes
    parameters are integrated in the application to HL~Tau ($i = 46.7^\circ$).
    Middle panel: gas density and amplitudes of the magnetic field components as
    functions of projected position ($l$) along the highlighted (thick red)
    field line in the left panel.
    The density is normalized by its mid-plane value whereas the field
    amplitudes are scaled by the maximum value attained by the dominant ($B_r$)
    component.
    Right panel: predicted polarization map, after convolution with the depicted
    synthesized beam, for a disc viewed at $i = 46.7^\circ$.
    The lengths of the plotted $E$-vectors scale with the degree of polarization
    $P$, which is calculated under the assumption that the ratio
    $C/C_\mathrm{I}$ in equation~(\ref{eq:P}) is a spatial constant.
    \label{fig:disk_wind}}
\end{figure*}
Figure~\ref{fig:3Ddisk} (3D rendering) and the left panel of
Fig.~\ref{fig:disk_wind} (2D vertical slice) show the distribution of the gas
mass density $\rho$ and the magnetic field geometry in our model.
The distributions of $\rho$ and of the three components of the magnetic field
$\mathbfit{B}$ along a representative field line are plotted in the middle panel
of Fig.~\ref{fig:disk_wind}.
It is seen that, in this model, the amplitude of the radial field component
increases rapidly with height above the mid-plane (where $B_r = B_\phi = 0$),
and that, over most of the distance between the mid-plane and the disc surfaces,
$|B_r| > |B_\phi| > |B_z|$.
In particular, near the disc surfaces, $|B_r/B_z|$ is well in excess of the
threshold for the launching of a centrifugally driven wind
(see Section~\ref{sec:intro}).
In the outflow region above the disc, all field amplitudes decrease with
distance along the field line.
The behaviour of the ratio $|B_r/B_\phi|$ is similar to the one obtained in
semi-analytic models of laminar, strongly coupled wind-driving discs in both the
(anisotropic) ambipolar and (isotropic) Ohm diffusivity regimes, for which
$|dB_r/dB_\phi|_0 = 2\Lambda_0$ \citep[e.g.][]{Konigl+10}.

For the post-processing, we truncate the disc at an outer radius of $175\,$au
and remove the central region (delimited by the inner dashed lines in the left
panel of Fig.~\ref{fig:disk_wind}) to avoid integrating through the inner
boundary of the simulation.
Adopting the disc inclination angle inferred for HL~Tau ($i = 46.7^\circ$;
\citealt{ALMA+15}), we generate synthetic polarization maps by considering a
virtual detector in the ($\tilde{x}$,~$\tilde{y}$) plane that consists of
$16\times16$ pixels, each associated with a distinct line of sight that
intersects the disc--wind structure (dotted lines in the left panel of
Fig.~\ref{fig:disk_wind}).
Although the 1.3-mm image of the HL~Tau disc exhibits a pattern of bright and
dark rings, with the latter possibly corresponding to physical gaps, we
approximate both the gas and the dust distributions as being smooth, with the
dust particle density $n_\mathrm{d}$ scaling as $\rho$.
Furthermore, even though the emission from the bright rings is inferred to be at
least marginally optically thick at this wavelength \citep{Pinte+16, Jin+16}, we
simplify the analysis by calculating the Stokes parameters in the optically thin
limit. Thus we write
\begin{equation}
  I = C_\mathrm{I}\int B_\nu n_\mathrm{d}\,\mathrm{d}s\,,\quad
  \left\{\begin{gathered}
  Q = C\int B_\nu n_\mathrm{d}\cos2\psi\cos^2\zeta\,\mathrm{d}s \\
  U = C\int B_\nu n_\mathrm{d}\sin2\psi\cos^2\zeta\,\mathrm{d}s
  \end{gathered}\right.
  \label{eq:stokes}
\end{equation}
\cite[e.g.][]{WardleKonigl90}, where the integrals are performed along each of
the chosen rays.
In equation~(\ref{eq:stokes}), $B_\nu(T_\mathrm{d})$ is the Planck function and
$T_\mathrm{d}$ the dust temperature, $\psi$ is the angle between the projected
magnetic field vector and the $\tilde{y}$ axis (measured counter-clockwise),
$\zeta$ is the angle between $\mathbfit{B}$ and its projection on the detector
plane, and $C$, $C_\mathrm{I}$ are constants.
The quantities within the integrals are all functions of $s$, with $\mathrm{d}s$
being the line element along the given ray.

For the dust temperature, we follow \citet{Stephens+14} and use the model fit of
\citet{Kwon+11} to the mid-plane and surface disc temperatures (subscripts 0 and
s, respectively) in HL~Tau as functions of the spherical radius $R$:
$T_0 = 190\,(R\,[\mathrm{au}])^{-0.43}\,$K,
$T_\mathrm{s} = 600\,(R\,[\mathrm{au}])^{-0.43}\,$K.
The disc surfaces are taken to be flat, and to be inclined at an angle of
$6^\circ$ to the equatorial plane (black solid lines in the left panel of
Fig.~\ref{fig:disk_wind}): these are approximately the loci where the radial
velocity changes sign, marking the transition from accretion to outflow.
Within the disc the dust temperature is obtained by interpolation, whereas for
the wind (subscript w) we assume a higher value, $T_\mathrm{w} = 3T_\mathrm{s}$
\citep[e.g.][]{BansKonigl12}.
While this temperature distribution favours emission from high latitudes, our
assumption that $n_\mathrm{d} \propto \rho$ implies -- in view of the strong
density stratification (middle panel of Fig.~\ref{fig:disk_wind}) -- that the
dominant contribution to the total intensity $I$ comes from the mid-plane
region.\footnote{
For $\lambda = 1.25$\,mm, $B_\nu(T_\mathrm{d})n_\mathrm{d}$ is approximately
proportional to $T_\mathrm{d}n_\mathrm{d}$.
Since the temperature only changes by a factor of a few over several scale
heights, whereas the density exhibits an exponential behaviour, it is
$n_\mathrm{d}$ that determines the emission.
This expectation is confirmed by our results, which are insensitive to the
particular choice of $T_\mathrm{w}$.}
However, as we discuss in Section~\ref{sec:discussion}, additional factors may
affect the distribution of both the polarized and the total dust emissivities in
the disc.

Once the Stokes parameters are calculated, we perform a Gaussian convolution
($Q\to Q_\mathrm{conv}$, $U\to U_\mathrm{conv}$, $I\to I_\mathrm{conv}$) using a
virtual circular beam of size $0.6''$ ($84\,$au), similar to the one used by
\citet{Stephens+14} in HL~Tau ($0.65''\times0.56''$).
The polarization position angle $\chi$ and the degree of linear polarization $P$
are then obtained from
\begin{equation}
  \tan2\chi = \frac{U_\mathrm{conv}}{Q_\mathrm{conv}}\ ,\quad
  P = \frac{C}{C_\mathrm{I}}\frac{\left(Q_\mathrm{conv}^2
    + U_\mathrm{conv}^2\right)^{1/2}}{I_\mathrm{conv}}\ .
  \label{eq:P}
\end{equation}
The right panel of Fig.~\ref{fig:disk_wind} shows the projected electric field
vectors for our disc model.\footnote{
The projected magnetic field vectors -- hereafter referred to as $B$-vectors --
are obtained by rotating the electric polarization vectors -- hereafter referred
to as $E$-vectors -- by $90^\circ$.
The lengths of the displayed polarization vectors are proportional to the
magnitude of $P$ at the projected locations of their centres.}
We treat the ratio $C/C_\mathrm{I}$ as a spatial constant in view of the fact
that the shapes and alignment properties of the radiating grains, which
determine this ratio, are uncertain.
We do, however, take into account the effect on the value of $P$ of
cancellations that arise either from the integration along the line of sight or
from the beam convolution procedure.
Both $\chi$ and $P$ vary with position in the disc, reflecting their dependence
on the projected field geometry (quantified by the angles $\psi$ and $\zeta$;
note that only the component of $\mathbfit{B}$ that is normal to the line of
sight contributes to the polarization) and on the projected distance from the
centre of the disc (as the effect of beam averaging is stronger near the
centre).

\section{Results}
  \label{sec:results}

\begin{figure}
  \includegraphics[width=\columnwidth]{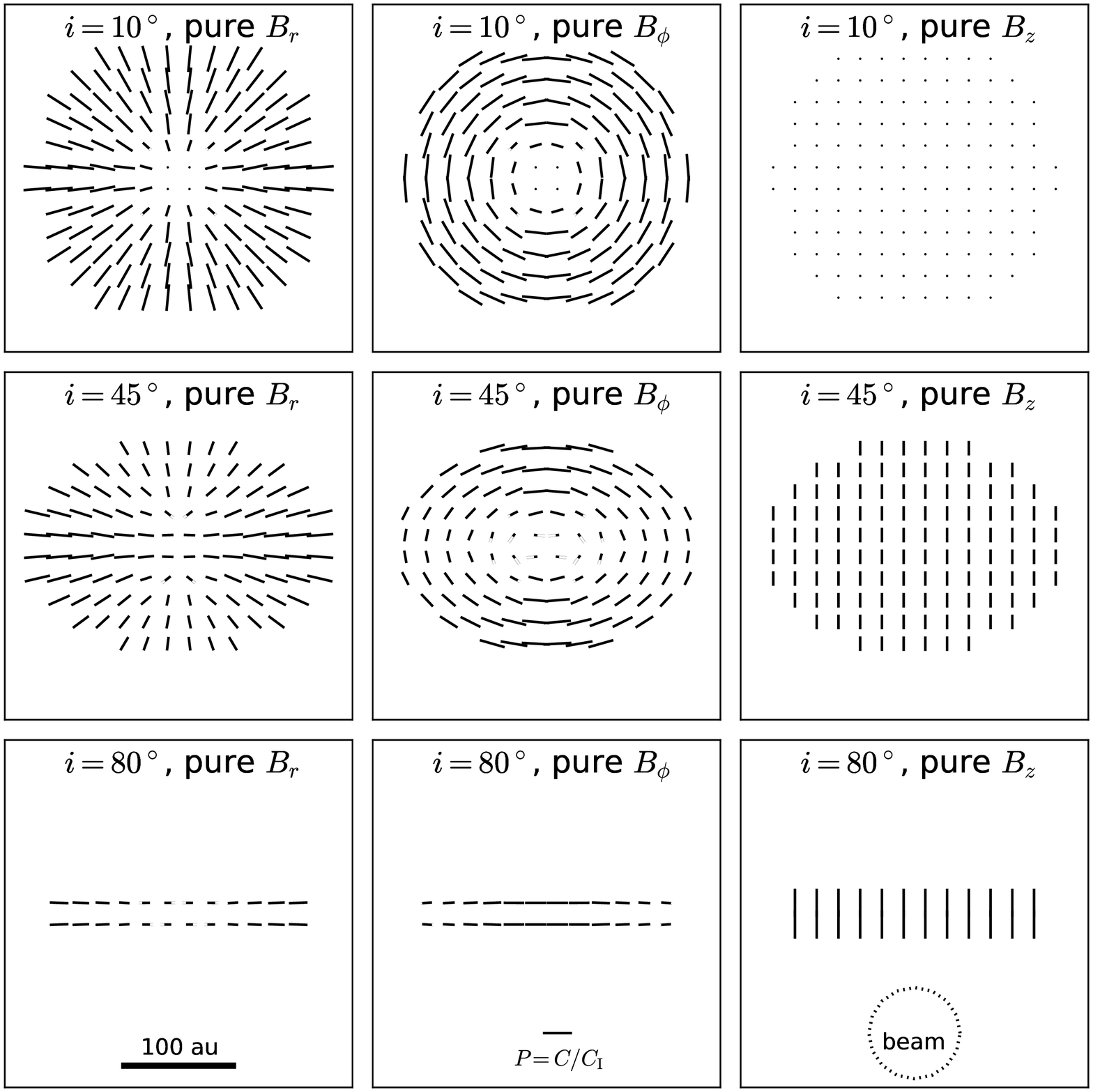}
  \caption{
    Heuristic $B$-vector maps obtained in the same manner as the $E$-vector map
    presented in the right panel of Fig.~\ref{fig:disk_wind}, except that the
    simulated magnetic field is replaced at each point by a single field
    component: radial (left panels), azimuthal (middle panels) or vertical
    (right panels).
    Each of these idealized configurations is shown for three viewing angles:
    $i=10^\circ$, $45^\circ$ and $80^\circ$ (top to bottom).
    Note that the joining of the $B$-vectors across the mid-plane in the lower
    right panel is an artefact of the limited resolution of the figure.
    \label{fig:Bcases}}
\end{figure}
Before turning to the specific predictions of our model, we present in
Fig.~\ref{fig:Bcases} synthetic polarization maps for three idealized magnetic
field configurations -- purely radial ($B_r$, left panels), purely azimuthal
($B_\phi$, middle panels) and purely vertical ($B_z$, right panels), each shown
for three different disc viewing angles -- nearly face on ($i=10^\circ$), nearly
edge on ($i=80^\circ$) and at an intermediate inclination ($i=45^\circ$; see
also \citealt{Aitken+02}).
It is seen that most of the reference configurations exhibit clearly
distinguishable polarization patterns.
The sole exception arises in the case of edge-on systems, where the basic
pattern is the same for both radial and azimuthal field geometries (with the
projected magnetic field direction being parallel to the plane of the disc).
However, this degeneracy is broken when one takes into account the degree of
polarization: in the pure $B_\phi$ case $P$ peaks at the centre of the projected
disc, whereas $P$ is largest near the edges when the field is radial.
Another manifestation of this behaviour is seen at intermediate viewing angles:
for the pure-$B_\phi$ case the degree of polarization is maximized on the
projected minor axis, whereas for a purely radial field the strongest
polarization occurs on the major axis.
These trends are readily understood from the expected emission properties of
aligned oblate grains \citep[see, e.g.][]{Yang+16b}.

\begin{figure*}
  \includegraphics[width=\textwidth]{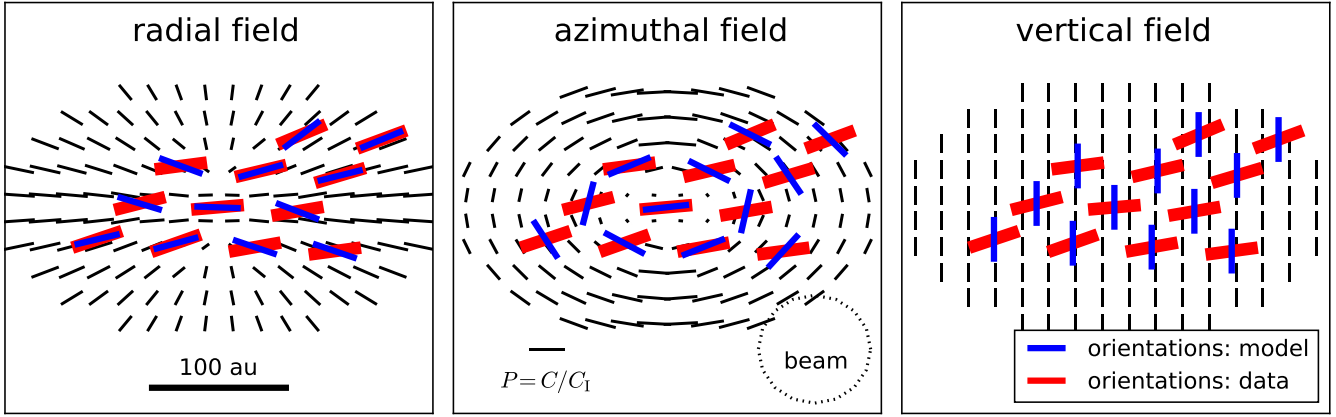}
  \caption{
    Synthetic polarization maps for the three magnetic field configurations
    considered in Fig.~\ref{fig:Bcases}, plotted using the disc viewing angle
    inferred in HL~Tau ($i = 46.7^\circ$).
    Superposed on the convolved $B$-vector distributions (black) are the
    $B$-vector orientation segments presented by \citeauthor{Stephens+14}
    (\citeyear{Stephens+14}; red) and the corresponding model predictions
    (blue).
    \label{fig:BcasesHLTau}}
\end{figure*}
The same three reference field configurations are shown in
Fig.~\ref{fig:BcasesHLTau} for the inferred inclination of the HL~Tau disc.
In each panel we compare the field orientations inferred from the $>3\,\sigma_P$
data \citep[red segments, taken from fig.~1 of][]{Stephens+14} with the
corresponding model predictions (blue segments).\footnote{
The observed position angle of the HL~Tau disc is $138^\circ$
\citep[measured counter-clockwise from north;][]{ALMA+15}; therefore, to match
the horizontal orientation of the major axis of our model disc, we rotated the
displayed data clockwise by $48^\circ$.}
This comparison suggests that a primarily vertical field is inconsistent with
the observations throughout the disc. Moreover, even though an azimuthal field
could in principle be compatible with the data in the central region, it is not
possible to match the observed orientations at the edges of the disc's major
axis.
These inferences are in agreement with the conclusions of \cite{Stephens+14},
who found that neither of these two field geometries, nor any combination
thereof, provides a good fit to the observations.
In contrast, the polarization pattern associated with a radial magnetic field
component (left panel) -- not explored by \cite{Stephens+14} -- appears to be in
much better agreement with the data.
The observations also indicate a higher degree of polarization away from the
disc's minor axis \citep[see fig.~1 of][]{Stephens+14}, which provides further
support to the inference of an underlying radial field. 

\begin{figure*}
  \includegraphics[width=\textwidth]{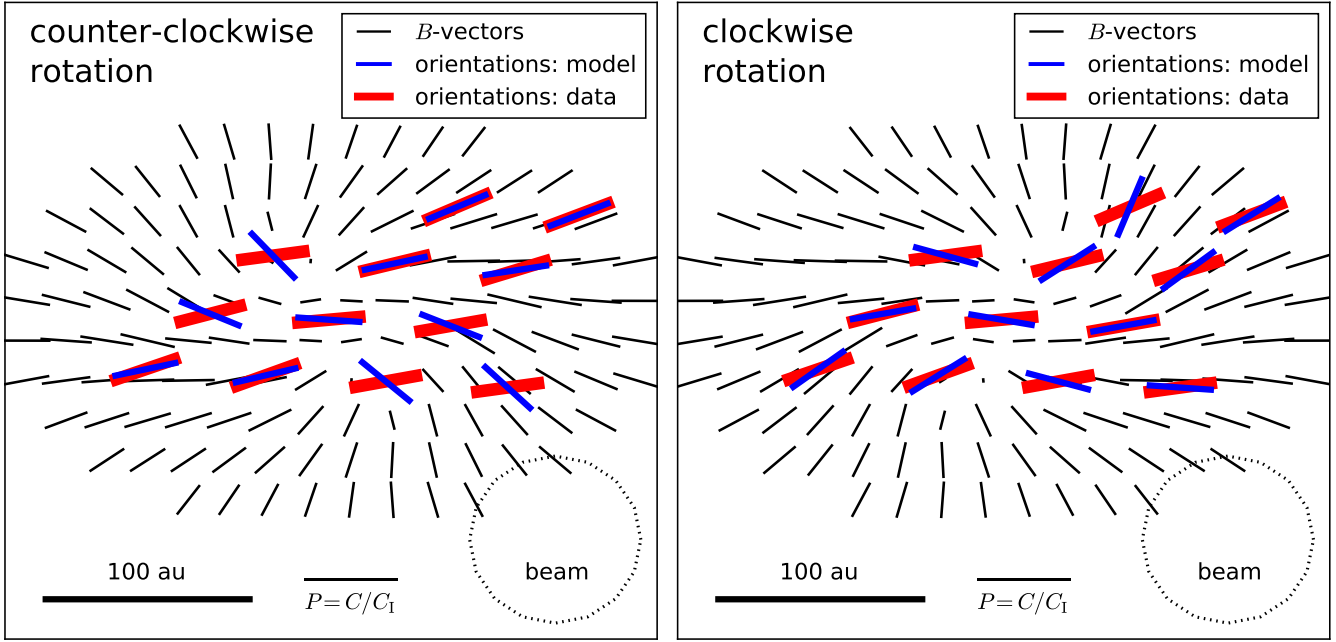}
  \caption{
    $B$-vector maps predicted by the simulated wind-driving disc model for
    counter-clockwise (left) and clockwise (right) disc rotations.
    The comparison with the data for HL~Tau is presented as in
    Fig.~\ref{fig:BcasesHLTau}.
    \label{fig:magnetic_field}}
\end{figure*}
The model described in Section~\ref{sec:model} represents a physical realization
of a disc in which the radial magnetic field component is dominant.
Figure~\ref{fig:magnetic_field} shows the predicted polarization pattern for
this model, with the actual data for HL~Tau again presented for comparison.
As expected on the basis of the preceding discussion, this model provides a
fairly good match to the observations.
To quantify the goodness of fit, we evaluated the quantity
$\epsilon \equiv (\mathrm{RSS}/N)^{1/2}$, where $N = 12$ is the number of data
points and RSS is the sum of the squares of the differences between the angular
coordinates of the model and the data $B$-vectors: the smaller the value of
$\epsilon$, the better the fit.\footnote{
Only the position-angle error ($\pm4.4^\circ$) of the central $B$-vector is
reported in \citet{Stephens+14}, so we cannot calculate other goodness-of-fit
measures, such as the reduced chi-square.}
Applying this measure to the reference cases shown in
Fig.~\ref{fig:BcasesHLTau}, we find $\epsilon_\mathrm{vertical} = 1.32$,
$\epsilon_\mathrm{azimuthal} = 0.92$ and $\epsilon_\mathrm{radial} = 0.33$,
which is consistent with the conclusions we reached on the basis of a visual
inspection of the figure.
In the case of the wind-driving disc model that we constructed, the presence of
a non-negligible azimuthal field component in addition to the (dominant) radial
magnetic field endows the polarization pattern with a weak curl whose sense
depends on the direction of the disc's rotation.
In the absence of additional information about the system, both orientations are
possible (corresponding to the projected disc being viewed from either its top
or its bottom side), and thus we present both options in the two panels of
Fig.~\ref{fig:magnetic_field}.
In principle, one could use the polarization data to infer the sense of rotation
of the underlying disc.
In this case, it is not entirely obvious which of the two models provides a
better fit: while the clockwise option appears to match a larger number of data
points, the counter-clockwise model -- even as it matches fewer points -- seems
to provide a better fit to the ones that it does match.
Using our chosen estimator, we find
$\epsilon_\mathrm{counter\mbox{-}clockwise} = 0.54$ and
$\epsilon_\mathrm{clockwise} = 0.34$, which favours clockwise rotation.
However, by combining the information on the disc kinematics provided by the
HCO$^+$ channel maps shown in fig.~4 of \citet{ALMA+15} with the [SII] data on
the radial velocities in the HL~Tau jets shown in fig.~8 of \citet{Mundt+90}, we
deduce that the HL~Tau disc in fact rotates counter-clockwise.
This suggests that our simple estimator is not sensitive enough to pick up the
correct sense of disc rotation in this source.
However, the $\epsilon$ value assigned in this scheme to the counter-clockwise
model still clearly favours it over models that lack a radial magnetic field
component.\footnote{
Note that an inherent property of our model is that the relative signs of $B_r$
and $B_\phi$ at each point are compatible with a magnetic stress that brakes the
disc's rotation.
\citet{WardleKonigl90} used a toy model of a geometrically thin disc that is
threaded by a magnetic field possessing an even symmetry with respect to the
mid-plane to infer this property directly from the polarization data for the
Galactic-centre disc.}

\begin{figure*}
  \includegraphics[width=\textwidth]{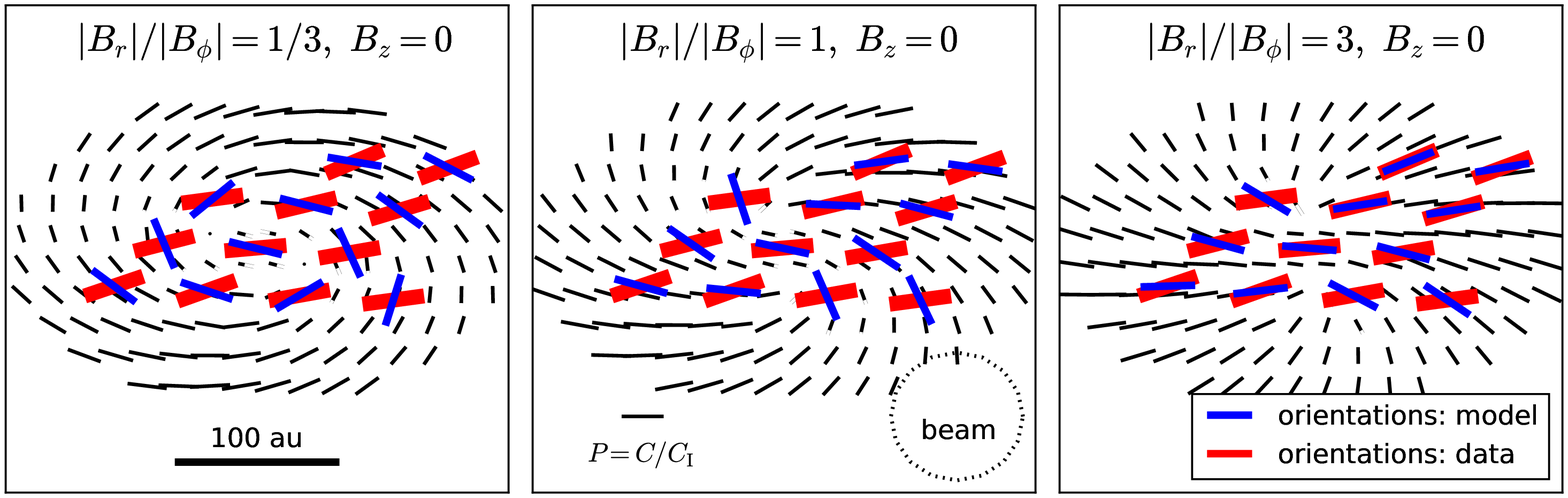}
  \caption{
    Synthetic polarization maps for three combinations of radial and azimuthal
    field components, presented in the same format as Fig.~\ref{fig:BcasesHLTau}
    under the assumption of counter-clockwise disc rotation.}
    \label{fig:radial_toroidal}
\end{figure*}
While the turbulent magnetic field produced by the MRI mechanism is
predominantly azimuthal, it also contains a significant radial component, with
the ratio between the mean radial and azimuthal field amplitudes being
$\sim$1/3 \citep[e.g.][]{Hawley+95, SanoStone02}.
For comparison, in our numerical disc--wind model, the ratio $|B_r|/|B_\phi|$
along a field line lies in the range $\sim$2--4 (with most of the polarized
emission originating near the upper end of this range).
To compare the predictions of these scenarios in relation to the HL~Tau data, we
employ our heuristic model to generate synthetic polarization maps for three
combinations of radial and azimuthal magnetic fields -- corresponding to
$|B_r|/|B_\phi| = 1/3$, $1$ and $3$ -- in a counter-clockwise rotating disc (see
Fig.~\ref{fig:radial_toroidal}).
The $\epsilon$ values obtained for these configurations are
$\epsilon_{1/3} = 0.88$, $\epsilon_1 = 0.83$ and $\epsilon_3 = 0.42$.
These results indicate that a good fit to the data requires
$|B_r|/|B_\phi| \gtrsim 3$, so that, in particular, an MRI disc model cannot
account for the observations.

\section{Discussion}
  \label{sec:discussion}

The numerical model we employed in this paper captures the essential ingredients
of a wind-driving disc: sufficiently good field--matter coupling (expressed by
the requirement that the Elsasser number $\Lambda$ be $>1$) for the field to
develop a radial component, and a sufficiently large field amplitude
(corresponding, in the ambipolar diffusivity regime, to the condition
$\beta \la 4\Lambda$; see \citealt{Salmeron+07}) for the magnetic field to be
strong enough to drive a CDW.
We used this model as a ``proof of concept'' but did not attempt to optimize the
fit by fine-tuning the parameters: this will be done more profitably when data
of higher resolution and S/N become available.
In addition, real protoplanetary discs are unlikely to have $\Lambda_0 > 1$ and
$\beta_0 \ga 1$ at all radial locations as assumed in our model.
In general, well-coupled and strongly magnetized models of the type that we
employed predict a mean accretion speed that is not much smaller than the local
speed of sound, which, in turn, implies a rather short mass depletion time-scale
for the disc.
Furthermore, there already exists direct evidence for relatively low interior
ionization levels in protoplanetary discs on $\ga 10\,$au scales \citep[e.g.][]
{Cleeves+15}, indicating that good coupling may be present only near the disc
surfaces.
As we noted in Section~\ref{sec:intro}, several numerical models of
$\beta_0 \gg 1$ discs that drive winds from comparatively high elevations have
already appeared in the literature.
However, these models are not yet fully global and thus cannot be used to
generate synthetic polarization maps.

As we also already noted in Section~\ref{sec:intro}, the high-$\beta_0$ disc
models predict that the azimuthal field component should dominate in the
vicinity of the mid-plane, implying -- if the dust density traces the gas
density -- that the polarization pattern in HL~Tau would resemble that shown in
the central panel of Fig.~\ref{fig:BcasesHLTau} rather than the much better
fitting $B_r$-dominated patterns shown in Fig.~\ref{fig:magnetic_field}.
In fact, in this source there is direct observational evidence for dust settling
to the mid-plane, with the scale height of the mm-wavelength dust disc inferred
(from the sharpness of the bright rings in the ALMA image) to be just
$\sim$1\,au at a distance of 100\,au \citep{Pinte+16}.
To be sure, a predominantly azimuthal field geometry is consistent with the
polarization maps obtained in some of the other observed sources.
One example is the edge-on system L1527, in which \citet{Segura-Cox+15} inferred
this geometry from the orientation of the $E$-vectors (perpendicular to the
projected disc plane).
This inference is supported by the fact that the measured polarization is
strongest near the centre of the disc, which serves to distinguish this
configuration from that of a predominantly radial field (see bottom row of
Fig.~\ref{fig:Bcases}).
\citet{Cox+15} similarly inferred an azimuthal field geometry in IRAS~4A based
on the observed polarization pattern in this source (which, according to
\citealt{Yang+16b}, is viewed at $i\approx 35^\circ$).
Indeed, the observed pattern is similar to the one shown in the central panel of
Fig.~\ref{fig:Bcases}, and, in further accord with the synthetic map, the
measured degree of polarization peaks along the disc's projected minor axis
(most noticeably at $8.1\,$mm).
However, both of these sources are Class-0 objects, in which the observations
cannot distinguish between the disc and the inner molecular-cloud envelope.
This ambiguity is most pronounced in the case of the third Class-0 source
observed to date, the face-on system IRAS~16293-2242B.
\citet{Rao+14} identified a spiral magnetic field pattern in the polarization
map of this source, which they related to the morphology obtained in a
simulation of the collapse of a rotating, magnetized, molecular-cloud core
\citep{Padovani+12}.
However, in view of the results presented in Fig.~\ref{fig:magnetic_field}, it
might be possible to interpret this pattern alternatively in terms of a
wind-driving disc.

Returning to the case of HL~Tau, where the possible contribution of an infalling
envelope is not an issue, how can one reconcile the strong indication of a
dominant radial field component in the polarization map with the expectation
that the bulk of the mm-wavelength emission originates near the disc mid-plane,
where the azimuthal field component dominates?
One possibility is that the non-negligible optical depth inferred in the bright
emission rings of HL~Tau at mm wavelengths \citep{Pinte+16, Jin+16} shifts the
emission centroid to finite disc elevations where the magnetic field already has
a measurable radial component.
However, in view of the very small scale height inferred for the mm-emitting
dust in this source, this is unlikely to be the main explanation.
Perhaps a more likely possibility is that, even in this comparatively young
source, the grains near the mid-plane, which dominate the total intensity, have
already grown to sizes that exceed the maximum size
$a_\mathrm{max} = \lambda/2\pi$ for producing polarized emission at wavelength
$\lambda$ (e.g. \citealt{ChoLazarian07}; for $\lambda = 1.25\,$mm,
$a_\mathrm{max}=0.2\,$mm), while the smaller grains (with sizes
$a < a_\mathrm{max}$), which contribute efficiently to the polarized flux,
remain suspended at high elevations (where the field is predominantly radial).
Another effect that could lower the polarized emission from grains that have
settled to the mid-pane is the likelihood that grains become less elongated as
they grow \citep[e.g.][]{Hughes+09}, which would tend to reduce the value of the
coefficient $C$ in equation~(\ref{eq:P}) ($C\rightarrow 0$ as the grain axis
ratio $\rightarrow 1$).\footnote{
A concentration of large grains near the mid-plane would also help explain the
sharp contrast between the bright and dark rings in this source \citep[see][]
{Ruge+16}.}
This interpretation is supported by the finding in the high-resolution
observations of IRAS~4A \citep{Cox+15} of an average polarization of 15\% at
$8\,$mm and 10\% at $10\,$mm, with a peak fractional polarization of $\sim$20\%.
If the intrinsic degree of mm-wavelength polarization in HL~Tau is also of the
order of 20\% then it may be possible to explain the factor of $\sim$10 lower
value of $P$ measured in this source at $1.25\,$mm\footnote{
\citet{Stephens+14} found that $P$ varies across the HL~Tau disc between
$0.54\pm0.13$\% and $2.4\pm 0.7$\%, with an average value of 0.90\%.}
in terms of a dilution of the polarized emission from $a \la 0.1\,$mm grains at
high disc elevations by weakly polarized emission of larger grains residing near
the mid-plane.
In this scenario, most of the grains that are responsible for the mm-wavelength
flux have settled to the mid-plane and grown to sizes $a \ga 1\,$mm.\footnote{
Note that \citet{Cox+15} concluded, based on their inferred upper limit
($\sim$3.2) on the dust spectral index in IRAS~4A, that a large number of
mm/cm-size grains are already present in this Class-0 system \citep[see, e.g.][]
{Draine06}.
This would suggest that grains with sizes that are at least as big -- if not
bigger -- could form in more evolved (Class-I/II) systems such as HL~Tau.
The spectral properties of the HL~Tau disc are consistent with this picture
\citep[e.g.][]{Kwon+11, Pinte+16, Carrasco-Gonzalez+16}.}
Although a fraction of these grains may have sizes in excess of $1\,$mm and
would therefore emit less efficiently at that wavelength than $a \la 1\,$mm
grains \citep[e.g.][]{MiyakeNakagawa93}, the mid-plane region should still
dominate the total mm-wavelength flux if most of the $a \ga 1\,$mm grains are
concentrated there.
Grains of size $a \la 0.1\,$mm may be kept at high elevations by turbulent
motions that can persist below the wind-driving surface layers \citep[e.g.][]
{Simon+13, Simon+15, Bai15} as well as by the emerging outflows within these
layers \citep[see][]{Safier93}, with porosity effects \citep[e.g.][]{Ormel+07}
possibly also helping to mitigate gravity's pull toward the mid-plane.
This scenario of course needs to be backed by detailed calculations and
observational tests.
One such test would be to obtain a polarization map of HL~Tau at longer
($\ga 1\,$cm) wavelengths: if the above picture is correct and the grains in the
mid-plane region are aligned, such a map could reveal a stronger (or even
dominant) contribution from the azimuthal and (especially if $\Lambda_0 \ll 1$)
vertical field components.\footnote{
A 7.0-mm JVLA image of the HL~Tau disc with a spatial resolution comparable to
that of the 2.9-mm ALMA image \citep{ALMA+15} was recently obtained by
\citet{Carrasco-Gonzalez+16}.
The 7.0-mm dust emission was inferred to be optically thin at all radii.}
It is, however, conceivable that the large grains in this source are not well
aligned because the radiative torque mechanism does not operate efficiently on
them: this could happen if the characteristic wavelength of the anisotropic
component of the local radiation field were much smaller than the mid-plane
grain sizes \citep[e.g.][]{ChoLazarian07} or if the anisotropic radiation
component inside the dust disc were weak due to finite optical-depth effects.
Note that the possibility of the mid-plane grains not being well aligned
provides another reason for why the polarized mm-wavelength emission from this
region could be weak.

A potential caveat to our proposed interpretation arises from the published
$\Lambda_0 < 1$, $\beta_0 \gg 1$ numerical disc models cited above, which
generally find $|B_\phi|/|B_r| > 1$ in the wind-launching region.
A similar behavior is exhibited by the previously obtained $\Lambda_0 < 1$,
$\beta_0 \ga 1$ semi-analytic solutions \citep{Li96,Wardle97}, and in both cases
it can be understood from the fact that, in the ambipolar diffusion-dominated
regime that applies on the $\ga 30$\,au radial scales of interest,
$|dB_r/dB_\phi|_0 = 2\Lambda_0$ (see Section~\ref{sec:model}).
However, as we have already emphasized, the existing models are not yet fully
global, which makes the locally inferred value of $|B_r|/|B_\phi|$ at the base
of the wind rather uncertain.
Indeed, the magnitudes of the radial and azimuthal field components at the top
of the disc are determined by constraints imposed \emph{outside} the disc
\citep[e.g.][]{OgilvieLivio01}; in particular, the value of $B_r$ at the base of
the wind is related to the magnetic flux distribution interior to the given
radial location, whereas the surface value of $B_\phi$ (which, for a given value
of $B_z$, specifies the magnetic torque that the wind exerts on the disc) is
related to the dynamical constraints (imposed at critical points) that a
steady-state wind solution must satisfy.
It is thus conceivable that a comprehensive, global disc--wind model can be
characterized by $|B_r|/|B_\phi| > 1$ at the base of the wind even if this ratio
is found to be $< 1$ in a local approximation.
As we inferred in Section~\ref{sec:results}, this ratio only needs to exceed
$\sim$3 to produce a good fit to the polarization data in HL~Tau.

An alternative interpretation of the HL~Tau polarization maps in terms of dust
scattering of the ambient disc radiation was considered by \citet{Kataoka+15,
Kataoka+16} and \citet{Yang+16a, Yang+16b}.
This mechanism has several distinctive features, although the details of the
predicted polarization patterns depend on the shapes, sizes and radiative
properties of the grains.
In the case of spherical grains, which only contribute to the polarized emission
through scattering, \citet{Yang+16a} showed that the predicted polarization
structure strongly resembles the one we found in this paper to be associated
with emission from aligned grains in a radial magnetic field.
In particular, the $E$-vectors -- which, in a disc viewed face on, trace
concentric circles -- are oriented predominantly along the disc's projected
minor axis (with $P$ peaking along the major axis) when $i=45^\circ$.
However, \citet{Kataoka+16} and \citet{Yang+16a} determined that the $\sim$1\%
mm-wavelength polarization measured in this source must originate in grains that
are much smaller ($a \la 100\,\micron$) than the mm/cm-size ones inferred to
produce the bulk of the emission.
This led \citet{Yang+16a} to a suggestion that is very similar to the one that
we proposed -- on different grounds -- in the context of our scenario, namely,
that the grains responsible for the polarized emission and the ones from which
most of the unpolarized emission originates represent two distinct populations,
with the large grains likely concentrated near the mid-plane and the smaller
ones probably floating at higher elevations.
However, most interstellar dust grains are evidently non-spherical (as first
indicated by observations of polarized starlight), so they can become
magnetically aligned and produce polarized emission.
\citet{Yang+16b} considered both the scattered- and the direct-emission
contributions and concluded that, on the spatial scales probed by
\citet{Stephens+14}, the polarized flux in a source like HL~Tau should be
dominated by the direct emission component.

Given that purely scattering (spherical) grains produce a polarization pattern
that closely resembles that from the direct emission of aligned grains in a
radial magnetic field, it could be anticipated that the joint contributions of
scattering and direct emission from non-spherical grains in an azimuthal
magnetic field might be able to mimic the pattern we obtain in our wind-driving
disc model, in which the field has both a radial and an azimuthal component.
In particular, it could be expected that this combination would be able to
reproduce the spiral magnetic field pattern that is generic to our model and is
most pronounced in our synthetic face-on disc maps (see the right panel of
Fig.~\ref{fig:disk_wind}).
However, as \citet{Yang+16b} pointed out, an oblate grain that is aligned with
its minor axis along an azimuthal field will generate a radial $E$-vector
pattern (rather than the azimuthal pattern produced by a spherical grain) when
located in the nearly isotropic radiation field that characterizes the inner
regions of the disc.
Since the direct emission component dominates in the outer disc, the predicted
polarization pattern for this case is radial (the same as the direct emission
component in an azimuthal field) on all scales.
For larger viewing angles, the predicted $E$-vectors fan out in a ``butterfly''
pattern, but, again, no spiral structure is produced.
This suggests that, if a spiral $B$-vector pattern such as the one identified in
IRAS 16293-2242B \citep{Rao+14} can be unambiguously associated with a
rotationally supported protoplanetary disc, this could serve as a clear
discriminant between the two models.

\section{Conclusion}
\label{sec:conclusion}

High-resolution mm- and cm-wavelength polarization maps of protoplanetary discs
on scales of $\sim$$10^2\,$au have recently been obtained for the Class-I/II
object HL~Tau and for several Class-0 protostars.
These maps hold the potential of probing the magnetic field structure and the
grain distribution in these discs, as revealed by the thermal emission from
aligned dust grains.
One of the main results of these observations has been the apparent absence of a
significant vertical magnetic field component, which was interpreted as evidence
against centrifugally driven winds (CDWs) playing a major role in the angular
momentum transport in these discs \citep[e.g.][]{Stephens+14, Segura-Cox+15}.
In this paper we demonstrate, focusing on the case of HL~Tau (where, in contrast
with the Class-0 systems, there is no doubt that the emission originates
exclusively in the disc), that this conclusion need not be true.
Using heuristic maps of the polarization patterns produced by
single-field-component (radial, azimuthal or vertical) discs as well as
synthetic maps generated by post-processing a numerical simulation of a
wind-driving disc, we show that the inclusion of a radial field component --
which was neglected in the published analyses -- significantly improves the fits
obtained by using only the azimuthal and vertical field components in modelling
the HL~Tau data.
In fact, we find that the best fit corresponds to a predominantly radial field.
Since a strong $B_r$ component is required for launching a CDW and is thus a
hallmark of wind-driving discs, we argue that, rather than providing evidence
against vertical magnetic angular momentum transport in protoplanetary discs,
the polarization maps may in fact corroborate that CDWs are present in these
systems.
A similar conclusion was previously reached in interpreting the far-infrared
polarization map of the pc-scale dust ring in the Galactic centre
\citep{Hildebrand+90, Hildebrand+93, WardleKonigl90}.

The numerical model that we use to demonstrate the basic polarization properties
of a wind-driving disc represents a system in which the mid-plane
magnetic-to-thermal pressure ratio is relatively large (corresponding to
$\beta_0$ being not much larger than~1), and in which the field--matter coupling
is strong even at $z=0$ (corresponding to the mid-plane Elsasser number,
$\Lambda_0$, being $>1$).
However, real discs may be characterized by $\beta_0 \gg 1$ and possibly also
$\Lambda_0 < 1$.
In this parameter regime, which numerical simulations have only recently started
to explore, disc outflows emerge only above a few thermal-pressure scale
heights, and the mid-plane region (where the bulk of the thermal dust emission
likely originates) is dominated by the $B_\phi$ component.
To reconcile this structure with our inference of a dominant $B_r$ component in
the 1.25-mm polarization map of HL~Tau, we suggest that most of the polarized
flux originates in $\la 0.1\,$mm-size grains that remain suspended above the
mid-plane and produce highly polarized ($P\ga 10\%$) emission, and that the
grains in the mid-plane region are larger and produce weakly polarized emission
at mm wavelengths.
(The implicit assumption underlying this proposal -- that the $B_r$ component
dominates in the surface regions of $\ga 30$\,au protoplanetary discs -- needs
to be verified using global disc--wind models.)
One possible (albeit inconclusive) test of this scenario is to search for
polarized emission from this source at longer wavelengths: if the grains at the
mid-plane region are well aligned, the maps would reveal stronger contributions
from the non-radial ($B_\phi$ and/or $B_z$) field components.

A recently proposed alternative model \citep{Kataoka+15, Kataoka+16, Yang+16a,
Yang+16b} attributes the polarization pattern observed in HL~Tau to dust
scattering of the disc radiation.
To be consistent with the spectral emission properties of this source, this
model also requires two different grain populations -- smaller ones at high
elevations and larger ones near the mid-plane -- to produce the bulk of the
polarized and the total fluxes, respectively.
One way to discriminate between the two models would be through the detection of
a spiral polarization pattern, like the one identified by \citet{Rao+14} in the
Class-0 source IRAS 16293-2242B (where, however, it could not be unambiguously
associated with the disc), in a post-Class-0 disc that is viewed at a relatively
small angle: such a pattern arises naturally only in the wind-driving disc
model.

\section*{Acknowledgements}

We are grateful to Ilsedore Cleeves, Mario Flock, Lynne Hillenbrand, Pat Palmer
and John Tobin for fruitful discussions.
We also thank the anonymous referee for several very useful comments.
This  work was supported in part by NASA ATP grant NNX13AH56G and completed with
resources provided by the University of Chicago Research Computing Center.
This research has made use of NASA's Astrophysics Data System Bibliographic
Services and of \texttt{matplotlib}, an open-source plotting library for Python
\citep{Hunter07}.

\bibliographystyle{mnras}
\bibliography{paper}

\begin{thebibliography}{}
\makeatletter
\relax
\def\mn@urlcharsother{\let\do\@makeother \do\$\do\&\do\#\do\^\do\_\do\%\do\~}
\def\mn@doi{\begingroup\mn@urlcharsother \@ifnextchar [ {\mn@doi@}
  {\mn@doi@[]}}
\def\mn@doi@[#1]#2{\def\@tempa{#1}\ifx\@tempa\@empty \href
  {http://dx.doi.org/#2} {doi:#2}\else \href {http://dx.doi.org/#2} {#1}\fi
  \endgroup}
\def\mn@eprint#1#2{\mn@eprint@#1:#2::\@nil}
\def\mn@eprint@arXiv#1{\href {http://arxiv.org/abs/#1} {{\tt arXiv:#1}}}
\def\mn@eprint@dblp#1{\href {http://dblp.uni-trier.de/rec/bibtex/#1.xml}
  {dblp:#1}}
\def\mn@eprint@#1:#2:#3:#4\@nil{\def\@tempa {#1}\def\@tempb {#2}\def\@tempc
  {#3}\ifx \@tempc \@empty \let \@tempc \@tempb \let \@tempb \@tempa \fi \ifx
  \@tempb \@empty \def\@tempb {arXiv}\fi \@ifundefined
  {mn@eprint@\@tempb}{\@tempb:\@tempc}{\expandafter \expandafter \csname
  mn@eprint@\@tempb\endcsname \expandafter{\@tempc}}}

\bibitem[\protect\citeauthoryear{{ALMA Partnership: Brogan} et~al.,}{{ALMA
  Partnership: Brogan} et~al.}{2015}]{ALMA+15}
{ALMA Partnership: Brogan} et~al., 2015, \mn@doi [\apjl]
  {10.1088/2041-8205/808/1/L3}, \href
  {http://adsabs.harvard.edu/abs/2015ApJ...808L...3P} {808, L3}

\bibitem[\protect\citeauthoryear{{Aitken}, {Efstathiou}, {McCall}  \&
  {Hough}}{{Aitken} et~al.}{2002}]{Aitken+02}
{Aitken} D.~K.,  {Efstathiou} A.,  {McCall} A.,   {Hough} J.~H.,  2002, \mn@doi
  [\mnras] {10.1046/j.1365-8711.2002.05047.x}, \href
  {http://adsabs.harvard.edu/abs/2002MNRAS.329..647A} {329, 647}

\bibitem[\protect\citeauthoryear{{Bai}}{{Bai}}{2013}]{Bai13}
{Bai} X.-N.,  2013, \mn@doi [\apj] {10.1088/0004-637X/772/2/96}, \href
  {http://adsabs.harvard.edu/abs/2013ApJ...772...96B} {772, 96}

\bibitem[\protect\citeauthoryear{{Bai}}{{Bai}}{2015}]{Bai15}
{Bai} X.-N.,  2015, \mn@doi [\apj] {10.1088/0004-637X/798/2/84}, \href
  {http://adsabs.harvard.edu/abs/2015ApJ...798...84B} {798, 84}

\bibitem[\protect\citeauthoryear{{Bai} \& {Stone}}{{Bai} \&
  {Stone}}{2013}]{BaiStone13}
{Bai} X.-N.,  {Stone} J.~M.,  2013, \mn@doi [\apj]
  {10.1088/0004-637X/769/1/76}, \href
  {http://adsabs.harvard.edu/abs/2013ApJ...769...76B} {769, 76}

\bibitem[\protect\citeauthoryear{{Balbus}}{{Balbus}}{2011}]{Balbus11}
{Balbus} S.~A.,  2011, in {Garcia} P.~J.~V.,  ed., Physical Processes in
  Circumstellar Disks around Young Stars. pp 237--282

\bibitem[\protect\citeauthoryear{{Bans} \& {K{\"o}nigl}}{{Bans} \&
  {K{\"o}nigl}}{2012}]{BansKonigl12}
{Bans} A.,  {K{\"o}nigl} A.,  2012, \mn@doi [\apj]
  {10.1088/0004-637X/758/2/100}, \href
  {http://adsabs.harvard.edu/abs/2012ApJ...758..100B} {758, 100}

\bibitem[\protect\citeauthoryear{{Blandford} \& {Payne}}{{Blandford} \&
  {Payne}}{1982}]{BlandfordPayne82}
{Blandford} R.~D.,  {Payne} D.~G.,  1982, \mnras, \href
  {http://adsabs.harvard.edu/abs/1982MNRAS.199..883B} {199, 883}

\bibitem[\protect\citeauthoryear{{Carrasco-Gonz{\'a}lez}
  et~al.,}{{Carrasco-Gonz{\'a}lez} et~al.}{2016}]{Carrasco-Gonzalez+16}
{Carrasco-Gonz{\'a}lez} C.,  et~al., 2016, \mn@doi [\apjl]
  {10.3847/2041-8205/821/1/L16}, \href
  {http://adsabs.harvard.edu/abs/2016ApJ...821L..16C} {821, L16}

\bibitem[\protect\citeauthoryear{{Cho} \& {Lazarian}}{{Cho} \&
  {Lazarian}}{2007}]{ChoLazarian07}
{Cho} J.,  {Lazarian} A.,  2007, \mn@doi [\apj] {10.1086/521805}, \href
  {http://adsabs.harvard.edu/abs/2007ApJ...669.1085C} {669, 1085}

\bibitem[\protect\citeauthoryear{{Cleeves}, {Bergin}, {Qi}, {Adams}  \&
  {{\"O}berg}}{{Cleeves} et~al.}{2015}]{Cleeves+15}
{Cleeves} L.~I.,  {Bergin} E.~A.,  {Qi} C.,  {Adams} F.~C.,   {{\"O}berg}
  K.~I.,  2015, \mn@doi [\apj] {10.1088/0004-637X/799/2/204}, \href
  {http://adsabs.harvard.edu/abs/2015ApJ...799..204C} {799, 204}

\bibitem[\protect\citeauthoryear{{Cox} et~al.,}{{Cox} et~al.}{2015}]{Cox+15}
{Cox} E.~G.,  et~al., 2015, \mn@doi [\apjl] {10.1088/2041-8205/814/2/L28},
  \href {http://adsabs.harvard.edu/abs/2015ApJ...814L..28C} {814, L28}

\bibitem[\protect\citeauthoryear{{Draine}}{{Draine}}{2006}]{Draine06}
{Draine} B.~T.,  2006, \mn@doi [\apj] {10.1086/498130}, \href
  {http://adsabs.harvard.edu/abs/2006ApJ...636.1114D} {636, 1114}

\bibitem[\protect\citeauthoryear{{Draine} \& {Weingartner}}{{Draine} \&
  {Weingartner}}{1996}]{DraineWeingartner96}
{Draine} B.~T.,  {Weingartner} J.~C.,  1996, \mn@doi [\apj] {10.1086/177887},
  \href {http://adsabs.harvard.edu/abs/1996ApJ...470..551D} {470, 551}

\bibitem[\protect\citeauthoryear{{Eardley} \& {Lightman}}{{Eardley} \&
  {Lightman}}{1975}]{EardleyLightman75}
{Eardley} D.~M.,  {Lightman} A.~P.,  1975, \mn@doi [\apj] {10.1086/153777},
  \href {http://adsabs.harvard.edu/abs/1975ApJ...200..187E} {200, 187}

\bibitem[\protect\citeauthoryear{{Ferreira}, {Dougados}  \&
  {Cabrit}}{{Ferreira} et~al.}{2006}]{Ferreira+06}
{Ferreira} J.,  {Dougados} C.,   {Cabrit} S.,  2006, \mn@doi [\aap]
  {10.1051/0004-6361:20054231}, \href
  {http://adsabs.harvard.edu/abs/2006A%26A...453..785F} {453, 785}

\bibitem[\protect\citeauthoryear{{Gressel}, {Turner}, {Nelson}  \&
  {McNally}}{{Gressel} et~al.}{2015}]{Gressel+15}
{Gressel} O.,  {Turner} N.~J.,  {Nelson} R.~P.,   {McNally} C.~P.,  2015,
  \mn@doi [\apj] {10.1088/0004-637X/801/2/84}, \href
  {http://adsabs.harvard.edu/abs/2015ApJ...801...84G} {801, 84}

\bibitem[\protect\citeauthoryear{{Hawley}, {Gammie}  \& {Balbus}}{{Hawley}
  et~al.}{1995}]{Hawley+95}
{Hawley} J.~F.,  {Gammie} C.~F.,   {Balbus} S.~A.,  1995, \mn@doi [\apj]
  {10.1086/175311}, \href {http://adsabs.harvard.edu/abs/1995ApJ...440..742H}
  {440, 742}

\bibitem[\protect\citeauthoryear{{Hennebelle} \& {Ciardi}}{{Hennebelle} \&
  {Ciardi}}{2009}]{HennebelleCiardi09}
{Hennebelle} P.,  {Ciardi} A.,  2009, \mn@doi [\aap]
  {10.1051/0004-6361/200913008}, \href
  {http://adsabs.harvard.edu/abs/2009A%26A...506L..29H} {506, L29}

\bibitem[\protect\citeauthoryear{{Hildebrand}, {Gonatas}, {Platt}, {Wu},
  {Davidson}, {Werner}, {Novak}  \& {Morris}}{{Hildebrand}
  et~al.}{1990}]{Hildebrand+90}
{Hildebrand} R.~H.,  {Gonatas} D.~P.,  {Platt} S.~R.,  {Wu} X.~D.,  {Davidson}
  J.~A.,  {Werner} M.~W.,  {Novak} G.,   {Morris} M.,  1990, \mn@doi [\apj]
  {10.1086/169248}, \href {http://adsabs.harvard.edu/abs/1990ApJ...362..114H}
  {362, 114}

\bibitem[\protect\citeauthoryear{{Hildebrand}, {Davidson}, {Dotson}, {Figer},
  {Novak}, {Platt}  \& {Tao}}{{Hildebrand} et~al.}{1993}]{Hildebrand+93}
{Hildebrand} R.~H.,  {Davidson} J.~A.,  {Dotson} J.,  {Figer} D.~F.,  {Novak}
  G.,  {Platt} S.~R.,   {Tao} L.,  1993, \mn@doi [\apj] {10.1086/173336}, \href
  {http://adsabs.harvard.edu/abs/1993ApJ...417..565H} {417, 565}

\bibitem[\protect\citeauthoryear{{Hughes}, {Wilner}, {Cho}, {Marrone},
  {Lazarian}, {Andrews}  \& {Rao}}{{Hughes} et~al.}{2009}]{Hughes+09}
{Hughes} A.~M.,  {Wilner} D.~J.,  {Cho} J.,  {Marrone} D.~P.,  {Lazarian} A.,
  {Andrews} S.~M.,   {Rao} R.,  2009, \mn@doi [\apj]
  {10.1088/0004-637X/704/2/1204}, \href
  {http://adsabs.harvard.edu/abs/2009ApJ...704.1204H} {704, 1204}

\bibitem[\protect\citeauthoryear{Hunter}{Hunter}{2007}]{Hunter07}
Hunter J.~D.,  2007, Computing In Science \& Engineering, 9, 90

\bibitem[\protect\citeauthoryear{{Jin}, {Li}, {Isella}, {Li}  \& {Ji}}{{Jin}
  et~al.}{2016}]{Jin+16}
{Jin} S.,  {Li} S.,  {Isella} A.,  {Li} H.,   {Ji} J.,  2016, preprint, \href
  {http://adsabs.harvard.edu/abs/2016arXiv160100358J} {} (\mn@eprint {arXiv}
  {1601.00358})

\bibitem[\protect\citeauthoryear{{Kataoka} et~al.,}{{Kataoka}
  et~al.}{2015}]{Kataoka+15}
{Kataoka} A.,  et~al., 2015, \mn@doi [\apj] {10.1088/0004-637X/809/1/78}, \href
  {http://adsabs.harvard.edu/abs/2015ApJ...809...78K} {809, 78}

\bibitem[\protect\citeauthoryear{{Kataoka}, {Muto}, {Momose}, {Tsukagoshi}  \&
  {Dullemond}}{{Kataoka} et~al.}{2016}]{Kataoka+16}
{Kataoka} A.,  {Muto} T.,  {Momose} M.,  {Tsukagoshi} T.,   {Dullemond} C.~P.,
  2016, \mn@doi [\apj] {10.3847/0004-637X/820/1/54}, \href
  {http://adsabs.harvard.edu/abs/2016ApJ...820...54K} {820, 54}

\bibitem[\protect\citeauthoryear{{Klaassen}, {Mottram}, {Maud}  \&
  {Juhasz}}{{Klaassen} et~al.}{2016}]{Klaassen+16}
{Klaassen} P.~D.,  {Mottram} J.~C.,  {Maud} L.~T.,   {Juhasz} A.,  2016,
  \mn@doi [\mnras] {10.1093/mnras/stw989}, \href
  {http://adsabs.harvard.edu/abs/2016MNRAS.tmp..826K} {}

\bibitem[\protect\citeauthoryear{{K{\"o}nigl} \& {Salmeron}}{{K{\"o}nigl} \&
  {Salmeron}}{2011}]{KoniglSalmeron11}
{K{\"o}nigl} A.,  {Salmeron} R.,  2011, in {Garcia} P.~J.~V.,  ed., Physical
  Processes in Circumstellar Disks around Young Stars. pp 283--352

\bibitem[\protect\citeauthoryear{{K{\"o}nigl}, {Salmeron}  \&
  {Wardle}}{{K{\"o}nigl} et~al.}{2010}]{Konigl+10}
{K{\"o}nigl} A.,  {Salmeron} R.,   {Wardle} M.,  2010, \mn@doi [\mnras]
  {10.1111/j.1365-2966.2009.15664.x}, \href
  {http://adsabs.harvard.edu/abs/2010MNRAS.401..479K} {401, 479}

\bibitem[\protect\citeauthoryear{{Kwon}, {Looney}  \& {Mundy}}{{Kwon}
  et~al.}{2011}]{Kwon+11}
{Kwon} W.,  {Looney} L.~W.,   {Mundy} L.~G.,  2011, \mn@doi [\apj]
  {10.1088/0004-637X/741/1/3}, \href
  {http://adsabs.harvard.edu/abs/2011ApJ...741....3K} {741, 3}

\bibitem[\protect\citeauthoryear{{Lesur}, {Kunz}  \& {Fromang}}{{Lesur}
  et~al.}{2014}]{Lesur+14}
{Lesur} G.,  {Kunz} M.~W.,   {Fromang} S.,  2014, \mn@doi [\aap]
  {10.1051/0004-6361/201423660}, \href
  {http://adsabs.harvard.edu/abs/2014A%26A...566A..56L} {566, A56}

\bibitem[\protect\citeauthoryear{{Li}}{{Li}}{1996}]{Li96}
{Li} Z.-Y.,  1996, \mn@doi [\apj] {10.1086/177469}, \href
  {http://adsabs.harvard.edu/abs/1996ApJ...465..855L} {465, 855}

\bibitem[\protect\citeauthoryear{{Looney}, {Mundy}  \& {Welch}}{{Looney}
  et~al.}{2000}]{Looney+00}
{Looney} L.~W.,  {Mundy} L.~G.,   {Welch} W.~J.,  2000, \mn@doi [\apj]
  {10.1086/308239}, \href {http://adsabs.harvard.edu/abs/2000ApJ...529..477L}
  {529, 477}

\bibitem[\protect\citeauthoryear{{Lumbreras} \& {Zapata}}{{Lumbreras} \&
  {Zapata}}{2014}]{LumbrerasZapata14}
{Lumbreras} A.~M.,  {Zapata} L.~A.,  2014, \mn@doi [\aj]
  {10.1088/0004-6256/147/4/72}, \href
  {http://adsabs.harvard.edu/abs/2014AJ....147...72L} {147, 72}

\bibitem[\protect\citeauthoryear{{Mignone}, {Bodo}, {Massaglia}, {Matsakos},
  {Tesileanu}, {Zanni}  \& {Ferrari}}{{Mignone} et~al.}{2007}]{Mignone+07}
{Mignone} A.,  {Bodo} G.,  {Massaglia} S.,  {Matsakos} T.,  {Tesileanu} O.,
  {Zanni} C.,   {Ferrari} A.,  2007, \mn@doi [\apjs] {10.1086/513316}, \href
  {http://adsabs.harvard.edu/abs/2007ApJS..170..228M} {170, 228}

\bibitem[\protect\citeauthoryear{{Mignone}, {Zanni}, {Tzeferacos}, {van
  Straalen}, {Colella}  \& {Bodo}}{{Mignone} et~al.}{2012}]{Mignone+12}
{Mignone} A.,  {Zanni} C.,  {Tzeferacos} P.,  {van Straalen} B.,  {Colella} P.,
    {Bodo} G.,  2012, \mn@doi [\apjs] {10.1088/0067-0049/198/1/7}, \href
  {http://adsabs.harvard.edu/abs/2012ApJS..198....7M} {198, 7}

\bibitem[\protect\citeauthoryear{{Miyake} \& {Nakagawa}}{{Miyake} \&
  {Nakagawa}}{1993}]{MiyakeNakagawa93}
{Miyake} K.,  {Nakagawa} Y.,  1993, \mn@doi [\icarus] {10.1006/icar.1993.1156},
  \href {http://adsabs.harvard.edu/abs/1993Icar..106...20M} {106, 20}

\bibitem[\protect\citeauthoryear{{Movsessian}, {Magakian}  \&
  {Moiseev}}{{Movsessian} et~al.}{2012}]{Movsessian+12}
{Movsessian} T.~A.,  {Magakian} T.~Y.,   {Moiseev} A.~V.,  2012, \mn@doi [\aap]
  {10.1051/0004-6361/201118529}, \href
  {http://adsabs.harvard.edu/abs/2012A%26A...541A..16M} {541, A16}

\bibitem[\protect\citeauthoryear{{Mundt}, {Buehrke}, {Solf}, {Ray}  \&
  {Raga}}{{Mundt} et~al.}{1990}]{Mundt+90}
{Mundt} R.,  {Buehrke} T.,  {Solf} J.,  {Ray} T.~P.,   {Raga} A.~C.,  1990,
  \aap, \href {http://adsabs.harvard.edu/abs/1990A%26A...232...37M} {232, 37}

\bibitem[\protect\citeauthoryear{{Ogilvie} \& {Livio}}{{Ogilvie} \&
  {Livio}}{2001}]{OgilvieLivio01}
{Ogilvie} G.~I.,  {Livio} M.,  2001, \mn@doi [\apj] {10.1086/320637}, \href
  {http://adsabs.harvard.edu/abs/2001ApJ...553..158O} {553, 158}

\bibitem[\protect\citeauthoryear{{Ormel}, {Spaans}  \& {Tielens}}{{Ormel}
  et~al.}{2007}]{Ormel+07}
{Ormel} C.~W.,  {Spaans} M.,   {Tielens} A.~G.~G.~M.,  2007, \mn@doi [\aap]
  {10.1051/0004-6361:20065949}, \href
  {http://adsabs.harvard.edu/abs/2007A%26A...461..215O} {461, 215}

\bibitem[\protect\citeauthoryear{{Padovani} et~al.,}{{Padovani}
  et~al.}{2012}]{Padovani+12}
{Padovani} M.,  et~al., 2012, \mn@doi [\aap] {10.1051/0004-6361/201219028},
  \href {http://adsabs.harvard.edu/abs/2012A%26A...543A..16P} {543, A16}

\bibitem[\protect\citeauthoryear{{Pinte}, {Dent}, {M{\'e}nard}, {Hales},
  {Hill}, {Cortes}  \& {de Gregorio-Monsalvo}}{{Pinte} et~al.}{2016}]{Pinte+16}
{Pinte} C.,  {Dent} W.~R.~F.,  {M{\'e}nard} F.,  {Hales} A.,  {Hill} T.,
  {Cortes} P.,   {de Gregorio-Monsalvo} I.,  2016, \mn@doi [\apj]
  {10.3847/0004-637X/816/1/25}, \href
  {http://adsabs.harvard.edu/abs/2016ApJ...816...25P} {816, 25}

\bibitem[\protect\citeauthoryear{{Rao}, {Girart}, {Lai}  \& {Marrone}}{{Rao}
  et~al.}{2014}]{Rao+14}
{Rao} R.,  {Girart} J.~M.,  {Lai} S.-P.,   {Marrone} D.~P.,  2014, \mn@doi
  [\apjl] {10.1088/2041-8205/780/1/L6}, \href
  {http://adsabs.harvard.edu/abs/2014ApJ...780L...6R} {780, L6}

\bibitem[\protect\citeauthoryear{{Ruge}, {Flock}, {Wolf}, {Dzyurkevich},
  {Fromang}, {Henning}, {Klahr}  \& {Meheut}}{{Ruge} et~al.}{2016}]{Ruge+16}
{Ruge} J.~P.,  {Flock} M.,  {Wolf} S.,  {Dzyurkevich} N.,  {Fromang} S.,
  {Henning} T.,  {Klahr} H.,   {Meheut} H.,  2016, \mn@doi [\aap]
  {10.1051/0004-6361/201526616}, \href
  {http://adsabs.harvard.edu/abs/2016A%26A...590A..17R} {590, A17}

\bibitem[\protect\citeauthoryear{{Safier}}{{Safier}}{1993}]{Safier93}
{Safier} P.~N.,  1993, \mn@doi [\apj] {10.1086/172574}, \href
  {http://adsabs.harvard.edu/abs/1993ApJ...408..115S} {408, 115}

\bibitem[\protect\citeauthoryear{{Salmeron}, {K{\"o}nigl}  \&
  {Wardle}}{{Salmeron} et~al.}{2007}]{Salmeron+07}
{Salmeron} R.,  {K{\"o}nigl} A.,   {Wardle} M.,  2007, \mn@doi [\mnras]
  {10.1111/j.1365-2966.2006.11277.x}, \href
  {http://adsabs.harvard.edu/abs/2007MNRAS.375..177S} {375, 177}

\bibitem[\protect\citeauthoryear{{Sano} \& {Stone}}{{Sano} \&
  {Stone}}{2002}]{SanoStone02}
{Sano} T.,  {Stone} J.~M.,  2002, \mn@doi [\apj] {10.1086/342172}, \href
  {http://adsabs.harvard.edu/abs/2002ApJ...577..534S} {577, 534}

\bibitem[\protect\citeauthoryear{{Segura-Cox}, {Looney}, {Stephens},
  {Fern{\'a}ndez-L{\'o}pez}, {Kwon}, {Tobin}, {Li}  \& {Crutcher}}{{Segura-Cox}
  et~al.}{2015}]{Segura-Cox+15}
{Segura-Cox} D.~M.,  {Looney} L.~W.,  {Stephens} I.~W.,
  {Fern{\'a}ndez-L{\'o}pez} M.,  {Kwon} W.,  {Tobin} J.~J.,  {Li} Z.-Y.,
  {Crutcher} R.,  2015, \mn@doi [\apjl] {10.1088/2041-8205/798/1/L2}, \href
  {http://adsabs.harvard.edu/abs/2015ApJ...798L...2S} {798, L2}

\bibitem[\protect\citeauthoryear{{Simon}, {Bai}, {Armitage}, {Stone}  \&
  {Beckwith}}{{Simon} et~al.}{2013}]{Simon+13}
{Simon} J.~B.,  {Bai} X.-N.,  {Armitage} P.~J.,  {Stone} J.~M.,   {Beckwith}
  K.,  2013, \mn@doi [\apj] {10.1088/0004-637X/775/1/73}, \href
  {http://adsabs.harvard.edu/abs/2013ApJ...775...73S} {775, 73}

\bibitem[\protect\citeauthoryear{{Simon}, {Lesur}, {Kunz}  \&
  {Armitage}}{{Simon} et~al.}{2015}]{Simon+15}
{Simon} J.~B.,  {Lesur} G.,  {Kunz} M.~W.,   {Armitage} P.~J.,  2015, \mn@doi
  [\mnras] {10.1093/mnras/stv2070}, \href
  {http://adsabs.harvard.edu/abs/2015MNRAS.454.1117S} {454, 1117}

\bibitem[\protect\citeauthoryear{{Stephens} et~al.,}{{Stephens}
  et~al.}{2014}]{Stephens+14}
{Stephens} I.~W.,  et~al., 2014, \mn@doi [\nat] {10.1038/nature13850}, \href
  {http://adsabs.harvard.edu/abs/2014Natur.514..597S} {514, 597}

\bibitem[\protect\citeauthoryear{{Takami}, {Beck}, {Pyo}, {McGregor}  \&
  {Davis}}{{Takami} et~al.}{2007}]{Takami+07}
{Takami} M.,  {Beck} T.~L.,  {Pyo} T.-S.,  {McGregor} P.,   {Davis} C.,  2007,
  \mn@doi [\apjl] {10.1086/524138}, \href
  {http://adsabs.harvard.edu/abs/2007ApJ...670L..33T} {670, L33}

\bibitem[\protect\citeauthoryear{{Tristram}, {Burtscher}, {Jaffe},
  {Meisenheimer}, {H{\"o}nig}, {Kishimoto}, {Schartmann}  \&
  {Weigelt}}{{Tristram} et~al.}{2014}]{Tristram+14}
{Tristram} K.~R.~W.,  {Burtscher} L.,  {Jaffe} W.,  {Meisenheimer} K.,
  {H{\"o}nig} S.~F.,  {Kishimoto} M.,  {Schartmann} M.,   {Weigelt} G.,  2014,
  \mn@doi [\aap] {10.1051/0004-6361/201322698}, \href
  {http://adsabs.harvard.edu/abs/2014A%26A...563A..82T} {563, A82}

\bibitem[\protect\citeauthoryear{{Tzeferacos}, {Ferrari}, {Mignone}, {Zanni},
  {Bodo}  \& {Massaglia}}{{Tzeferacos} et~al.}{2009}]{Tzeferacos+09}
{Tzeferacos} P.,  {Ferrari} A.,  {Mignone} A.,  {Zanni} C.,  {Bodo} G.,
  {Massaglia} S.,  2009, \mn@doi [\mnras] {10.1111/j.1365-2966.2009.15502.x},
  \href {http://adsabs.harvard.edu/abs/2009MNRAS.400..820T} {400, 820}

\bibitem[\protect\citeauthoryear{{Tzeferacos}, {Ferrari}, {Mignone}, {Zanni},
  {Bodo}  \& {Massaglia}}{{Tzeferacos} et~al.}{2013}]{Tzeferacos+13}
{Tzeferacos} P.,  {Ferrari} A.,  {Mignone} A.,  {Zanni} C.,  {Bodo} G.,
  {Massaglia} S.,  2013, \mn@doi [\mnras] {10.1093/mnras/sts266}, \href
  {http://adsabs.harvard.edu/abs/2013MNRAS.428.3151T} {428, 3151}

\bibitem[\protect\citeauthoryear{{Wardle}}{{Wardle}}{1997}]{Wardle97}
{Wardle} M.,  1997, in {Wickramasinghe} D.~T.,  {Bicknell} G.~V.,   {Ferrario}
  L.,  eds,  Astronomical Society of the Pacific Conference Series Vol. 121,
  IAU Colloq. 163: Accretion Phenomena and Related Outflows. p.~561 (\mn@eprint
  {} {astro-ph/9707228})

\bibitem[\protect\citeauthoryear{{Wardle} \& {Koenigl}}{{Wardle} \&
  {Koenigl}}{1993}]{WardleKonigl93}
{Wardle} M.,  {Koenigl} A.,  1993, \mn@doi [\apj] {10.1086/172739}, \href
  {http://adsabs.harvard.edu/abs/1993ApJ...410..218W} {410, 218}

\bibitem[\protect\citeauthoryear{{Wardle} \& {K\"onigl}}{{Wardle} \&
  {K\"onigl}}{1990}]{WardleKonigl90}
{Wardle} M.,  {K\"onigl} A.,  1990, \mn@doi [\apj] {10.1086/169249}, \href
  {http://adsabs.harvard.edu/abs/1990ApJ...362..120W} {362, 120}

\bibitem[\protect\citeauthoryear{{Yang}, {Li}, {Looney}  \& {Stephens}}{{Yang}
  et~al.}{2016a}]{Yang+16a}
{Yang} H.,  {Li} Z.-Y.,  {Looney} L.,   {Stephens} I.,  2016a, \mn@doi [\mnras]
  {10.1093/mnras/stv2633}, \href
  {http://adsabs.harvard.edu/abs/2016MNRAS.456.2794Y} {456, 2794}

\bibitem[\protect\citeauthoryear{{Yang}, {Li}, {Looney}, {Cox}, {Tobin},
  {Stephens}, {Segura-Cox}  \& {Harris}}{{Yang} et~al.}{2016b}]{Yang+16b}
{Yang} H.,  {Li} Z.-Y.,  {Looney} L.~W.,  {Cox} E.~G.,  {Tobin} J.,  {Stephens}
  I.~W.,  {Segura-Cox} D.~M.,   {Harris} R.~J.,  2016b, \mn@doi [\mnras]
  {10.1093/mnras/stw1253}, \href
  {http://adsabs.harvard.edu/abs/2016MNRAS.460.4109Y} {460, 4109}

\makeatother
\end{thebibliography}

\bsp
\label{lastpage}
\end{document}